\documentclass[a4paper,fleqn,english]{article}

 \usepackage{setspace,graphics}
 \usepackage[dvips]{epsfig} 
 \usepackage{a4,amssymb,epsfig,array,cite}

\usepackage{amsfonts,bm,amssymb,euscript,array,babel}

\renewcommand{\title}[1]{\vspace{10mm}\noindent{\Large{\bf #1}}\vspace{8mm}}
\newcommand{\authors}[1]{\noindent{\large #1}\vspace{5mm}}
\newcommand{\address}[1]{{\itshape #1\vspace{2mm}}}

\usepackage{graphicx}

\newcommand{\be}{\begin{equation}}
\newcommand{\ee}{\end{equation}}
\newcommand{\eq}[1]{(\ref{#1})}
\def\nn{\nonumber}
\def\bea{\begin{eqnarray}}
\def\eea{\end{eqnarray}}
\def\cM{{\cal M}}
\def\cC{{\cal C}}
\def\cA{{\cal A}}
\def\cH{{\cal H}}

\def\R{{\mathbb R}}

\def\Tr{{\rm Tr}}
\def\L{\Lambda}
\def\one{\mbox{1 \kern-.59em {\rm l}}}

\begin{document}



\sloppy

\begin{center}

\title{Emergent 4D Gravity from Matrix Models}

\authors{Harold Steinacker\footnote{supported by the FWF Project P18657}}

\address{Fakult\"at f\"ur Physik, Universit\"at Wien\\
 Boltzmanngasse 5, A-1090 Wien, Austria\\
E-mail:~\textsf{harold.steinacker@univie.ac.at}}

\begin{abstract}

Recent progress in the understanding of gravity 
on noncommutative spaces is discussed. 
A gravity theory naturally emerges from 
matrix models of noncommutative gauge theory.
The effective metric depends on the 
dynamical Poisson structure, absorbing the degrees of freedom 
of the would-be $U(1)$ gauge field. 
The gravity action is induced upon quantization.

\end{abstract}

\end{center}

\section{Background and motivation}

There is a fundamental conflict between quantum mechanics 
and general relativity at the Planck scale. 
This lead to the development of quantum  field theory on 
noncommutative spaces \cite{reviews}, as a first step to 
overcome this problem. 
More recently, it was suggested that gravity emerges naturally from 
noncommutative gauge theory, 
without having to introduce any new degrees of freedom such 
as an explicit metric. 
Earlier forms of this idea \cite{Yang:2006hj,Rivelles:2002ez}
can be cast in concise form for 
matrix models of noncommutative gauge theory \cite{Steinacker:2007dq},
which describe dynamical noncommutative spaces. 
We discuss basic results of this approach. This
also provides a new understanding of gravity in 
similar string theoretical matrix models \cite{Ishibashi:1996xs}.

\section{Matrix models and effective geometry}
\label{sec:metric}

Consider the matrix model with action 
\be
S_{YM} = - \Tr [Y^a,Y^b] [Y^{a'},Y^{b'}] g_{a a'} g_{b b'}
\label{YM-action-1}
\ee
for
\be
g_{a a'} = \delta_{a a'} \quad \mbox{or}\quad g_{a a'} = \eta_{a a'} 
\label{background-metric}
\ee
in the Euclidean  resp.  Minkowski case.
Here the "covariant coordinates"  $Y^a$ for $a=1,2,3,4$ 
are hermitian matrices 
or operators acting on some Hilbert space\footnote{the matrices 
are meant to be infinite-dimensional, but ``regularized'' by  
$N \times N$ matrices for $N \to \infty$.} 
$\cal{H}$.
We will denote their commutator as 
\be
[Y^a,Y^b] = i \theta^{ab}
\label{theta-def}
\ee
so that $\theta^{ab} \in L(\cal{H})$ is an 
antisymmetric matrix, which is 
{\em not} assumed to be constant here.
We focus on configurations $Y^a$ (not necessarily solutions of 
the equation of motion) which can be interpreted as quantizations 
of a Poisson manifold $(\cM,\theta^{ab}(y))$ with general
Poisson structure $\theta^{ab}(y)$. This defines the geometrical 
background under consideration, and conversely 
essentially any (local) Poisson manifold provides 
after quantization a possible background $Y^a$. 
More formally, this means that there is a map of vector spaces
(``quantization map'')
\be 
\cC(\cM) \to \cA \subset L(\cal{H})\, 
\label{map} 
\ee 
where $\cC(\cM)$ denotes the space of functions on $\cM$, 
and $\cA$ is interpreted as quantized algebra of 
functions on $\cM$. 
The map \eq{map} can be used to define a star product on $\cC(\cM)$.
Furthermore, we can then write
\be
\left[f,g\right] \sim i \{f(y),g(y)\} 
\ee
for $f,g \in \cA$,
where $\sim$ denotes the leading term in a semi-classical 
expansion in $\theta$,
and $\{f,g\}$ the Poisson bracket defined by $\theta^{ab}(y)$.
$Y^a$ can be interpreted as
quantization of a classical coordinate function 
$y^a$ on $\cM$. More importantly, $Y^a$ defines a derivation on 
$\cA$ via
\be
[Y^a,f] \sim i \theta^{ab}(y) \partial_b f(y) , \qquad f \in \cA .
\label{derivation}
\ee
Consider first the {\em ``irreducible'' case}
i.e. assume that the centralizer of $\cA$ in $\cal{H}$ is trivial. 
Then any matrix (``function'') in $L(\cH)$  
can be well approximated by a function of $Y^a$.
From the gauge theory point of view discussed in section 
\ref{sec:flat-expansion}, it
means that we restrict ourselves to the $U(1)$ case;
this is also the sector where the UV/IR mixing occurs. 
For the general case see section \ref{sec:nonabelian}.

Let us now couple a scalar field $\Phi \in {\cal A}$ 
to the matrix model \eq{YM-action-1}.
The only possibility 
to write down kinetic terms for matter fields is through commutators 
$[Y^a,\Phi] \sim i\theta^{ab}(y) \frac{\partial}{\partial y^b} \Phi$
using \eq{derivation}. This leads to the action
\be
S[\Phi] \, =\, - \Tr\, g_{ab} [Y^a,\Phi][Y^{b},\Phi]  
\, \sim\,  \int d^4 y\, \rho(y)\, 
G^{ab}(y)\,\frac{\partial}{\partial y^a}\Phi(y) 
\frac{\partial}{\partial y^b} \Phi(y) .
\label{scalar-action-0}
\ee
Here
\be
G^{ab}(y) = \theta^{ac}(y) \theta^{b d}(y)\, g_{cd} \, 
\label{effective-metric}
\ee
is the effective metric for the scalar field $\Phi$  
\cite{Steinacker:2007dq}.
Hence the Poisson manifold naturally 
acquires a metric structure $(\cM,\theta^{ab}(y),G^{ab}(y))$, 
which is determined by the Poisson structure and the 
constant background metric $g_{ab}$.
We also used $\Tr \sim \int  d^4 y\, \rho(y)$, where 
\be
\rho(y) = |\det G_{ab}(y)|^{1/4} = (\det\theta^{ab}(y))^{-1/2} 
\ee
is the symplectic measure on $(\cM,\theta^{ab}(y))$. 
Notice that the action \eq{scalar-action-0} is
invariant under Weyl rescaling of $\theta^{ab}(y)$ resp. $G^{ab}(y)$.
We can therefore write the action as
\be
S[\Phi] = \int d^4 y\, 
\tilde G^{ab}(y)\, \partial_{y^a}\Phi(y) \partial_{y^b}\Phi(y) 
= \int d^4 y\, \sqrt{|\tilde G_{ab}|}\,\, \Phi(y) \Delta_{\tilde G} \Phi(y) 
\label{scalar-action-geom}
\ee
where $\tilde G^{ab}$
is the unimodular metric 
\be
\tilde G^{ab}(y) = (\det G_{ab})^{1/4}\, G^{ab}(y),
\qquad \det \tilde G^{ab}(y) = 1
\label{metric-unimod}
\ee
and $\Delta_{\tilde G}$ is the Laplacian of a scalar field on
the classical Riemannian manifold $(\cM,\tilde G^{ab}(y))$.

The main point here is that any kinetic term will 
always involve the metric $G^{ab}(y)$ resp. $\tilde G^{ab}(y)$,
possibly with additional density factors which remain 
to be understood.
Therefore this metric
should indeed be interpreted as gravitational metric. 
For gauge fields this is discussed in section \ref{sec:nonabelian},
and the case of fermions will be discussed elsewhere.
Note also that $\theta^{ac}(y)$ 
can be interpreted as a preferred frame or vielbein, 
which is however gauge-fixed and does not admit the usual 
local Lorentz resp. orthogonal transformations.

A linearized version of \eq{metric-unimod} 
was obtained in \cite{Rivelles:2002ez}.
Related (but inequivalent) metrics were discussed
in the context of the DBI action \cite{Yang:2006hj};
note that our metric $G^{ab}$ which governs the matrix model
is {\em not} the pull-back
of $g^{ab}$ using the change of coordinates \eq{cov-coord-1},
and it is indeed curved in general.

It is easy to see that in 4 dimensions, 
one cannot obtain the most general 
geometry from metrics of the form \eq{effective-metric}.
Therefore the gravity theory which emerges here will not
reproduce all (off-shell) degrees of freedom of general relativity.
However, one does obtain a class of metrics
which is sufficiently rich to describe the propagating
(``on-shell'') degrees of freedom of gravity, 
as well as e.g. the Newtonian limit for an 
arbitrary mass distribution \cite{Steinacker:2007dq}. 
On noncommutative spaces,
the 2 physical helicities of gravitons can indeed be 
expressed in terms of the 2 physical helicities of photons.

\paragraph{Equations of motion.}

So far we considered arbitrary background configurations $Y^a$ as long
as they admit a geometric interpretation.
The equations of motion derived from the action \eq{YM-action-1} 
\be
[Y^a,[Y^{a'},Y^b]]\, g_{a a'} = 0 \, 
\label{eom}
\ee
select on-shell geometries among all possible backgrounds, such as 
the Moyal-Weyl quantum plane \eq{Moyal-Weyl}. 
These amount to Ricci-flat
spaces \eq{R-flat} at least in the near-flat case \cite{Rivelles:2002ez}. 
However we allow the 
most general off-shell configurations here.

\section{Quantization and induced gravity}

Now consider the quantization of the matrix model \eq{YM-action-1}
coupled to a scalar field.
In principle, the quantization is defined in terms of a 
(``path'') integral
over all matrices $Y^a$ and $\Phi$. 
In 4 dimensions, only perturbative computations
can be performed for the gauge sector encoded by $Y^a$, 
while the scalar $\Phi$ can be integrated
out formally in terms of a determinant. Let us focus here on the
effective action obtained by integrating out the scalar,
\be
e^{-\Gamma_{\Phi}} = \int d\Phi e^{-S[\Phi]} , \quad\mbox{where}\quad
 \Gamma_{\Phi} = \frac 12 \Tr \log \Delta_{\tilde G} 
\ee
for a non-interacting scalar field with action \eq{scalar-action-geom}.
A standard argument using the heat kernel expansion
of $\Delta_{\tilde G}$ leads to
\bea
\Gamma_{\Phi} &=& \frac 1{16\pi^2}\, \int d^4 y \,\left( -2\L^4 
+ \frac 16 R[\tilde G]\, \L^2 + O(\log \L) \right)\, \nn\\
 &=& \frac 1{16\pi^2}\, \int d^4 y \, \left(-2\L^4 + 
\frac 16 \rho(y)\left(R[G] -  3 \Delta_G \sigma-\frac 32\,G^{ab}
\partial_a\sigma \partial_b \sigma\right)\L^2 + O(\log \L) \right)\nn\\
\label{S-oneloop-scalar}
\eea
where
\bea
\Delta_G \sigma &=&  G^{ab} \partial_a \partial_b \sigma - \Gamma^c \partial_c \sigma , \qquad \Gamma^a = G^{bc}\, \Gamma_{bc}^a\, , \nn\\
e^{-\sigma(y)} &=& \rho(y) = (\det G_{ab})^{1/4}
\eea
This is essentially the mechanism of induced gravity 
\cite{Sakharov:1967pk}, and it suggests to identify the gravitational 
constant with the cutoff $\frac 1G \sim \L^2$.
Note that the term $\int d^4 y\sqrt{\tilde G}\, \L_y^4$
is usually interpreted as cosmological constant, and its 
scaling with $\L^4$ presents a major problem for induced
gravity. However  
$\det \tilde G =1$ here,  which suggests that this term
is essentially trivial in the present context.
One finds indeed that flat space \eq{Moyal-Weyl}
is a solution even at one loop, in sharp contrast with 
general relativity.
These are strong hints that the 
notorious cosmological constant problem is absent
or at least much milder in this NC gravity theory. 

A further very remarkable point is that the 
above gravitational effective action
provides an understanding of the UV/IR mixing 
in NC gauge theory:
\eq{S-oneloop-scalar} gives the physical content of
the ``strange'' IR behavior of NC gauge theory in a suitable regime.
This will be elaborated in detail elsewhere.

\section{Gauge theory point of view}

In this section we discuss the alternative 
(more conventional, up to now) interpretation of 
\eq{YM-action-1}, \eq{scalar-action-0}
in terms of NC gauge theory. This will also set the 
stage for the extension to $SU(n)$ gauge theory coupled
to  gravity.

\subsection{Geometry from $U(1)$ gauge fields}
\label{sec:flat-expansion}

Let us now rewrite the actions 
\eq{YM-action-1}, \eq{scalar-action-0} in terms of the $U(1)$ gauge fields on 
the flat Moyal-Weyl background $\R^4_\theta$ with generators 
$\bar X^a$. 
This means that we consider ``small fluctuations''
\be
Y^a = \bar X^{a} + \cA^a\, 
\label{cov-coord-1}
\ee
around the generators $\bar X^{a}$ of the 
Moyal-Weyl quantum plane, which satisfy
\be
[\bar X^a,\bar X^b] = i \bar\theta^{ab}\, .
\label{Moyal-Weyl}
\ee
Here $\bar\theta^{ab}$ 
is a constant antisymmetric non-degenerate tensor. 
More precisely, we assume that the hermitian matrices
$\cA^a =\cA^a(\bar X) \sim \cA^a(x)$ can be interpreted (at
least ``locally'') as smooth functions 
on $\R^4_{\bar \theta}$. 
Note that the effective geometry \eq{effective-metric}
 for the Moyal-Weyl plane is indeed
flat, given by
\bea
\bar g^{ab} &=& \bar\theta^{ac}\,\bar\theta^{bd} g_{cd}\, \nn\\
\tilde g^{ab} &=& \, \bar\rho\, \bar g^{ab},
\qquad \bar\rho = |\det \bar g_{ab}|^{1/4} 
= (\det\bar\theta^{ab})^{-1/2} \equiv \L_{NC}^4.
\label{effective-metric-bar}
\eea
Consider now the
change of variables
\be
\cA^a(x) = -\bar\theta^{ab} A_b(x)
\label{A-naive}
\ee
where $A_a$ is hermitian. 
Using
\be
[\bar X^a + \cA^a,f] = i \bar\theta^{ab} (\frac{\partial}{\partial x^b} f
+ i [A_b,f])  \equiv  i \bar\theta^{ab} D_b f,
\ee
the actions \eq{YM-action-1}, \eq{scalar-action-0} can be written as 
\bea
S[\Phi] &=& \Tr\,\bar\theta^{ab}\,\bar\theta^{a'c} g_{aa'}\,
 D_a \Phi\, D_b \Phi
 = \int d^4 x\, \tilde g^{ab}\,  D_a\Phi(x) D_b \Phi(x) , \nn\\
S_{YM} &=& \int d^4 x\,\bar \rho\, (\bar g^{a a'}\,\bar g^{b b'}\,F_{ab}\,F_{a'b'}
\, + \bar g^{ab} g_{ab})
\label{scalar-action-A-tilde}
\eea
where $F_{ab} = \partial_a A_b - \partial_a A_b  + i [A_a,A_b]\,$ is the
$U(1)$ field strength.
These formulas are exact (up to boundary terms) if interpreted as 
noncommutative gauge theory on $\R^4_{\bar \theta}$. 
In the geometrical
interpretation \eq{scalar-action-geom} the gauge field $U(1)$ gauge
field $A_a(x)$ is completely
absorbed in the metric $G^{ab}(y)$ resp. $\theta^{ab}(y)$.

\subsection{Nonabelian gauge fields}
\label{sec:nonabelian}

We now discuss the extension of the above model
to nonabelian gauge fields
\be
S_{YM} = - \Tr [X^a,X^b] [X^{a'},X^{b'}] g_{a a'} g_{b b'}\, .
\label{YM-action-nonabel}
\ee
The action is formally the same as \eq{YM-action-1}, but we use different
letters for the matrices 
hoping to avoid possible confusions.
Consider the new vacuum given by the reducible solution 
$X^a = \bar X^a\otimes \one_n$ of the equation of motion \eq{eom}.
The most general matrix near this vacuum can be written as 
\be
X^a = \bar X^a\otimes \one_n + \cA^a
   = Y^a \otimes \one_n + \cA^{a,\alpha}(Y)\otimes \lambda_\alpha
\label{covar-coord-nonabel}
\ee
where $Y^a = \bar X^a + \cA^{a,0}$ denotes the trace-$U(1)$ sector,
and $\lambda_\alpha$ the $SU(n)$ Gell-mann matrices.
It is well-known that this can be interpreted as 
$U(n)$ gauge fields on the Moyal-Weyl quantum plane;
in particular, $\cA^{a,0}$ is usually interpreted as $U(1)$ 
gauge field. 
However, generalizing the argument in section \ref{sec:metric}
it is more natural to absorb the $\cA^{a,0}$ in the geometry.
Indeed it was shown in \cite{Steinacker:2007dq} that
the model \eq{YM-action-nonabel} can be 
interpreted  as 
$SU(n)$ gauge fields coupled to gravity, with the same effective
metric \eq{effective-metric} as above. 
This explains why the $U(1)$ sector cannot be disentangled from the 
$SU(n)$ gauge fields in the noncommutative case.

Technically speaking, the analysis of the semiclassical limit
of \eq{YM-action-nonabel} requires the use of the 
Seiberg-Witten map \cite{Seiberg:1999vs} for general noncommutativity 
$\theta^{ab}(y)$. This allows to express the fluctuations
$\cA^{a,\alpha}(Y)$ through {\em commutative} $SU(n)$ gauge fields 
on $\cM$, and ensures that the resulting
semi-classical action is gauge invariant. The 
$SU(n)$ field strength is contracted with the effective metric
$G^{ab}(y)$ as expected, up to a density factor
(which remains to be understood).
One finds after considerable effort 
\cite{Steinacker:2007dq}
\be
S_{YM} \,\sim\, S_{\rm eff} =  \int d^4 y\, \rho(y) \,
tr \left(4\eta(y) + G^{a a'} G^{b b'} F_{ab}\, F_{a'b'}\right)
 - 2 \int \eta(y) \, tr F\wedge F 
\label{S-YM-effective}
\ee 
up to higher-oder corrections in $\theta$,
where $F_{ab}$ 
is the $SU(n)$ field strength on $\cM$, and
\be
\eta(y) = \frac 14 \,G^{ab}(y) g_{ab} \,.
\ee
Remarkably, this involves a ``would-be topological term'' 
$\int \eta(y) \, tr F\wedge F$. This may have implications
for the strong CP problem. Furthermore, note that $S_{\rm eff}$ 
appears to be generally covariant and 
invariant under local Lorentz transformations, if we consider
$\eta(y), \rho(y)$ as a scalar functions.
This is remarkable, because Lorentz-invariance 
would appear to be violated
from the Moyal-plane point of view.
However, these are not fundamental
symmetries because $g_{ab}$ is fixed in \eq{YM-action-nonabel}.
It is also remarkable that the (``would-be'' $U(1)$) 
term $\int d^4 y\, \rho(y) \eta(y)$ 
implies that the vacuum geometries are Ricci-flat, 
\be
R_{ab}[\tilde G] \, =\, 0 \quad + O(\theta^2)
\label{R-flat}
\ee
at least in linearized gravity 
$\tilde G_{ab} = \bar g_{ab} + h_{ab}$ \cite{Rivelles:2002ez}.
Finally, as shown in \cite{Steinacker:2007dq}
the class of metrics \eq{effective-metric}
is rich enough to describe correctly the
Newtonian limit of gravity, with metric
\be
ds^2 = -c^2 dt^2\Big(1+\frac {2U}{c^2}\Big) 
+  d \vec x^2 \Big(1+O(\frac 1{c^2})\Big) .
\label{newton-metric}
\ee
Here $\Delta_{(3)} U = 4\pi G\rho$ and $\rho$ is the mass density.
Combined with \eq{S-oneloop-scalar}, we see that a reasonable candidate
for physical gravity emerges.
It promises advantages over GR for quantization and the 
cosmological constant problem.
The most exciting aspect is that it provides an extremely
simple and intrinsically noncommutative mechanism for gravity.
On the other hand the constrained class 
of metrics \eq{effective-metric} makes the 
theory very restrictive (and falsifiable); adding
extra dimensions might extend this class of metrics.
Of course much more work
is required to obtain a complete understanding and judgement.




\begin{thebibliography}{[1]}




\bibitem{reviews} for basic reviews see
  M.~R.~Douglas and N.~A.~Nekrasov,
  ``Noncommutative field theory,''
  Rev.\ Mod.\ Phys.\  {\bf 73} (2001) 977
  [arXiv:hep-th/0106048];
  R.~J.~Szabo,
  ``Quantum field theory on noncommutative spaces,''
  Phys.\ Rept.\  {\bf 378}, 207 (2003)
  [arXiv:hep-th/0109162].


\bibitem{Rivelles:2002ez}
  V.~O.~Rivelles,
  ``Noncommutative field theories and gravity,''
  Phys.\ Lett.\  B {\bf 558} (2003) 191
  [arXiv:hep-th/0212262].


\bibitem{Yang:2006hj}
  H.~S.~Yang,
  ``On The Correspondence Between Noncommuative Field Theory And Gravity,''
  Mod.\ Phys.\ Lett.\  A {\bf 22} (2007) 1119;
  H.~S.~Yang,
  ``Emergent gravity from noncommutative spacetime,''
  arXiv:hep-th/0611174

\bibitem{Steinacker:2007dq}
  H.~Steinacker,
  ``Emergent Gravity from Noncommutative Gauge Theory,''
  JHEP {\bf 12}, (2007) 049; 
 [arXiv:0708.2426 [hep-th]].

\bibitem{Ishibashi:1996xs}
  N.~Ishibashi, H.~Kawai, Y.~Kitazawa and A.~Tsuchiya,
  ``A large-N reduced model as superstring,''
  Nucl.\ Phys.\  B {\bf 498} (1997) 467
  [arXiv:hep-th/9612115]



\bibitem{Seiberg:1999vs}
  N.~Seiberg and E.~Witten,
  ``String theory and noncommutative geometry,''
  JHEP {\bf 9909} (1999) 032
  [arXiv:hep-th/9908142].


\bibitem{Sakharov:1967pk}
  A.~D.~Sakharov,
  ``Vacuum quantum fluctuations in curved space and the theory of
  gravitation,''
  Sov.\ Phys.\ Dokl.\  {\bf 12} (1968) 1040
  [Dokl.\ Akad.\ Nauk Ser.\ Fiz.\  {\bf 177} (1967)].



\end{thebibliography}
\end{document}